\documentclass[12pt,a4paper]{article}
\usepackage[utf8]{inputenc}
\usepackage{amsmath}
\usepackage{amsfonts}
\usepackage{amssymb}
\usepackage{graphicx}
\usepackage{hyperref}
\newcommand\eq[1]{\begin{eqnarray}#1\end{eqnarray}}

\newcommand{\diff}{\mathrm{d}}
\newcommand{\dr}{\mathrm{d}r}
\newcommand{\du}{\mathrm{d}u}
\newcommand{\dv}{\mathrm{d}v}
\newcommand{\ad}{\mathrm{ad}}
\newcommand{\Ad}{\mathrm{Ad}}
\newcommand{\tr}{\mathrm{Tr}}

\author{Heikki Arponen}
\title{On generalizations of asymptotically $AdS_3$ spaces and geometry of $SL(N)$}
\begin{document}
\maketitle
\begin{abstract}
In three and two dimensions the asymptotic symmetry groups of $AdS$ spaces are infinite dimensional. This can be explained easily by noting the relations $AdS_3 \simeq SL(2)$ and $AdS_2 \simeq SL(2)/SO(2)$, i.e. that the asymptotic symmetries are in fact that of the Lie group $SL(2)$. As show in the author's previous work, similar infinite dimensional asymptotic symmetry groups can be found in the case of $SL(3)$ and probably also for other noncompact Lie groups and their homogeneous spaces.
The purpose of the present work is to revisit the $AdS_3$ space in detail from the Lie group point of view by finding the boundary theory energy-momentum tensor and to prepare to tackle the $SL(3)$ and $SL(N)$ cases.
\end{abstract}
\newpage

\section{Introduction}

The anti de-Sitter space in dimension $d+1$ can be defined as the homogeneous space $SO(2,d)/SO(2,d-1)$ with an isometry group $SO(2,d)$.\footnote{Every Lie group is here assumed to be defined over the field of real numbers.} In the low dimensional special cases the asymptotic symmetry groups (which preserve the asymptotically $AdS$ structure) can be shown to be the Virasoro algebra in $d=1$ and two copies of the Virasoro algebras in $d=2$, corresponding to boundary conformal transformations\cite{henneaux}. For $d>2$ the asymptotic symmetries form a finite dimensional Lie algebra, which is isomorphic to $so(2,d)$.\\

The anti de-Sitter space in three dimensions is particularly interesting. It was shown in \cite{henneaux} that an asymptotically $AdS_3$ gravity is dual to a two dimensional conformal field theory on the boundary of the $AdS$ space, which can be seen as an explicit manifestation of the holographic principle. It was later shown that this dual CFT is the 2D Liouville theory (see e.g. \cite{henneaux.coussaert}). It should be pointed out that these dualities do not require any sort of AdS/CFT prescription between bulk and boundary fields but occur quite naturally. In fact, it is possible to find explicitly the boundary CFT's energy-momentum tensor from the near horizon $AdS$ metric (see e.g. \cite{Nakatsu}), as will be show also in this work. The construction depends crucially on the infinite dimensionality of the asymptotic symmetry algebra. It is therefore not surprising that similar dualities cannot be found for the higher dimensional $AdS$ spaces, since their asymptotic symmetry algebras are too small.\\

It is instructive to investigate the low dimensional spaces in greater detail by noting the isomorphisms $SO(1,2) \simeq SL(2)$ and $SO(2,2) \simeq SL(2)\times SL(2)$. Then $AdS_2 \simeq SL(2)/SO(2)$ and $AdS_3 \simeq SL(2)$. The isometry groups can be understood as the Lie group $SL(2)$ acting on the left coset from the left in the case of $AdS_2$ and acting on the Lie group itself both from the left and right in the $AdS_3$ case. It is therefore evident that the infinite dimensional asymptotic symmetry algebras are strongly connected to the Lie group $SL(2)$. For $d>2$ no such relations to the special linear groups exist.\\

It is now tempting to ask whether such infinite dimensional asymptotic symmetries can occur in the case of other special linear Lie groups and their homogeneous spaces. Indeed, it was shown in \cite{arponen} that the asymptotic symmetries of the five dimensional homogeneous (symmetric) space $SL(3)/SO(3)$ do form an infinite dimensional Lie algebra. However, the analysis in that work was more of a brute force approach and only to leading order (\emph{at} the boundary). To be able to find the energy-momentum tensor, the analysis must be extended to subleading orders (\emph{near} boundary). This poses technical challenges that the naive approach of the aforementioned work will not be able to handle. The purpose of the present work is to present tools that will ease the work in the more difficult cases (particularly $SL(3)$) and to review the $AdS_3$ case using these tools.

\section{Asymptotic symmetries of $SL(2)$}
The formalism below applies in a more general setting than just $SL(2)$, but for the sake of a more gentle introduction, facts about $SL(2)$ and its Lie algebra are presented on the side. While all of the explicit results have already been know for quite some time, the reader is urged to keep in mind the possible generalization to e.g. $SL(3)$. Another key aspect is that the group theoretic formalism is more elegant and reveals some intricacies that are not easily observed in an approach where one simply tries to find asymptotic symmetries of a given metric or its fall-off conditions.

\subsection{The Lie algebra and Lie group}

The Lie algebra $sl(2;\mathbb R)$ will be defined here for simplicity as the 2*2 traceless matrices
\eq{X_1 = \frac{1}{2}\left( \begin{matrix}
0 & 1 \\ 
1 & 0
\end{matrix}\right), X_2 = \frac{1}{2}\left( \begin{matrix}
-1 & 0 \\ 
0 & 1
\end{matrix}\right), X_3 = \frac{1}{2}\left( \begin{matrix}
0 & 1 \\ 
-1 & 0
\end{matrix}\right) ,}
which satisfy the commutation relations
\eq{\left[ X_i, X_j \right] = {{\epsilon'}_{i j}}^k X_k}
with $1={{\epsilon'}_{1 2}}^3 = - {{\epsilon'}_{2 3}}^1 = -{{\epsilon'}_{3 1}}^2$. This is actually the Lie algebra $so(2,1)$, which however is isomorphic to $sl(2)$. We also define the flat metric and the inner product with the help of the Killing form as
\eq{\eta_{i j} \doteq 2 \tr \left( X_i X_j\right) \doteq \langle X_i , X_j \rangle. \label{innerproduct}}
then $\eta$ is simply the Minkowskian metric
\eq{\eta = \left( \begin{matrix}
1 & 0 & 0 \\ 
0 & 1 & 0 \\
0 & 0 & -1
\end{matrix}\right).}
It can be used to raise and lower indices, which results in particular that the permutation symbol is expressed as $\epsilon_{i j k} = {{\epsilon'}_{i j}}^l \eta_{l k}$. \\

The elements in the Lie group in the general $SL(N)$ case will be defined in terms of the Euler angle parametrization as $g=k h k'$, where $k$ and $k'$ are in the (different) compact subgroups $SO(N)$ and $h$ is in the Cartan subgroup. Explicitly in the $SL(2)$ case,
\eq{g = e^{u X_3} e^{r X_2} e^{v X_3}= e^{r/2}\left(
\begin{array}{cc}
 -\sin\left(\frac{u}{2}\right) \sin\left(\frac{v}{2}\right) & \cos\left(\frac{v}{2}\right) \sin\left(\frac{u}{2}\right) \\
 -\cos\left(\frac{u}{2}\right) \sin\left(\frac{v}{2}\right) & \cos\left(\frac{u}{2}\right) \cos\left(\frac{v}{2}\right)
\end{array}\right)\\   + e^{-r/2} \left(
\begin{array}{cc}
 \cos\left(\frac{u}{2}\right) \cos\left(\frac{v}{2}\right) & \cos\left(\frac{u}{2}\right) \sin\left(\frac{v}{2}\right) \\
 -\cos\left(\frac{v}{2}\right) \sin\left(\frac{u}{2}\right) & -\sin\left(\frac{u}{2}\right) \sin\left(\frac{v}{2}\right)
\end{array}
\right).}
\subsection{Maurer-Cartan forms and geometry}
The Lie algebra valued Maurer-Cartan (MC) forms are defined as $\Omega_g \doteq g^{-1} \diff g = X_i \otimes \Omega^i$ and $\overline \Omega_g \doteq  \diff g g^{-1}= X_i \otimes \overline \Omega^i$, respectively. Their one-form components are
\eq{\left\{ \begin{aligned}
\Omega^1 &= -\sin (v) \dr + \cos (v) \sinh (r) \du\\
\Omega^2 &= \cos(v) \dr +  \sin(v) \sinh (r) \du\\
\Omega^3 &= \cosh(r) \du + \dv
\end{aligned}
\right. }
and
\eq{\left\{ \begin{aligned}
\overline\Omega^1 &= \sin (u) \dr - \cos (u) \sinh (r) \dv\\
\overline\Omega^2 &= \cos(u) \dr +  \sin(u) \sinh (r) \dv\\
\overline\Omega^3 &= \du + \cosh(r) \dv
\end{aligned}
\right. .}
They will both naturally satisfy the Maurer-Cartan equations for a flat connection, 
\eq{\diff \Omega + \Omega \wedge \Omega=0.}
The inner product (\ref{innerproduct}) can be used to define the metric on the Lie group $SL(2;\mathbb R)$ as
\eq{&ds^2 \doteq \langle \Omega_g, \Omega_g \rangle \equiv \langle \overline\Omega_g, \overline\Omega_g \rangle\\
&= \dr^2 -\du^2 -\dv^2 -2 \du \dv \cosh(r).\label{metric}}
This is of course just the metric on $AdS_3$.
\subsection{Vector fields on $SL(2)$}
Denote the left and right actions on $G$ in the usual way as
\eq{\left\{ \begin{aligned}
& L_{ g'}: G \to G , L_{ g'} (g) \doteq  g' g\\
& R_{ g'}: G \to G , R_{ g'} (g) \doteq g  g' .\end{aligned}
\right.}
The left and right invariant vector fields acting on $g \in SL(2)$ are defined as
\eq{\widehat X (g) \doteq \left. \frac{\diff}{\diff \epsilon}\right\vert_{\epsilon=0}\!\!\!\!\!\!\!\! g e^{\epsilon X}= g X = L_g (X) }
for $X \in sl(2)$, where the "hat" is used to refer to a differential operator and similarly for the right invariant vector field.\footnote{E.g. the statement of left invariance means simply $(L_{g'})_* \widehat X (g) = g'g X = \widehat X (g'g)$.} Then for example
\eq{\widehat X_3 (g) = e^{u X_3} e^{r X_2} e^{v X_3} X_3 = \partial_v g
}
and similarly in the other cases. The vector fields read explicitly
\eq{\left\{ \begin{aligned}
\widehat X_1  &= -\sin(v) \partial_r + \frac{\cos(v)}{\sinh(r)} \partial_u - \cos(v) \coth(r) \partial_v\\
\widehat X_2  &= \cos(v) \partial_r + \frac{\sin(v)}{\sinh(r)} \partial_u - \sin(v) \coth(r) \partial_v\\
\widehat X_3  &= \partial_v
\end{aligned}
\right. \label{leftvectors}}
and
\eq{\left\{ \begin{aligned}
\widehat{\bar X}_1  &= \sin(u) \partial_r + \cos(u) \coth(r) \partial_u  -\frac{\cos(u)}{\sinh(r)} \partial_v \\
\widehat{\bar X}_2  &= \cos(u) \partial_r - \sin(u) \coth(r) \partial_u  +\frac{\sin(u)}{\sinh(r)} \partial_v\\
\widehat{\bar X}_3  &= \partial_u
\end{aligned}
\right. \label{rightvectors} }
The MC forms are dual to the vector fields, i.e. $\Omega^i (\widehat X_j) = \delta^i_j$. The left invariant vector fields satisfy the original commutation relations while the right invariant ones satisfy the negative of the original relations.
\subsection{Isometries and gauge transformations}
Suppose $\omega$ is a point in $G=SL(2)$. A gauge transformation $g'  \in SL(2)$, depending on the point $\omega$, transforms the MC forms as
\eq{\left\{ \begin{aligned}
\Omega'_g &\doteq \left( L_{ g'^{-1}}\right)^* \Omega_g = \Omega_{ g' g} = g^{-1} \Omega_{ g'} g + \Omega_g\\
\widetilde{\overline\Omega}_g &\doteq \left( L_{ g'^{-1}}\right)^* \overline\Omega_{g} = \overline\Omega_{ g' g} =  g' \overline\Omega_{ g}  g'^{-1} + \overline\Omega_{ g'}
\end{aligned}
\right. \label{gauge_finite} }
and similarly for the right actions. For a constant $ g'$ we have $\Omega_{\tilde g} =  \overline\Omega_{\tilde g}=0$ which results in $\Omega_g$ being invariant with respect to the transformation and $\overline\Omega'_g =  g' \overline\Omega_{ g} g'^{-1}$. Both of these transformations are isometries of the metric (\ref{metric}) due to the definition (\ref{innerproduct}) of the inner product. The isometries are naturally generated by the vector fields (\ref{leftvectors}) and respectively by (\ref{rightvectors}) for the right action. When ${g'} \approx 1 + \xi(\omega) \cdot X$ with $\xi(\omega)$ infinitesimally small, the change in the MC forms is
\eq{\left\{ \begin{aligned}
\delta \Omega_g &= \diff \xi^i \Ad^{-1}_{g}\left( X_i \right) = X_i {\left( \Ad^{-1}_{g} \right)^i}_j \diff \xi^j\\
\delta \overline \Omega_g &= \xi^i \ad_{X_i} \left( \overline\Omega_g \right) + \diff \xi^i X_i 
\end{aligned}
\right. \label{gauge_infinitesimal}}
where the adjoint action notations $\Ad_{g}\left( X \right) \doteq g X g^{-1}$ and $\ad_{X} \left( Y \right) \doteq [X,Y]$ were used. The vector field corresponding to the above gauge transformation is then just $\xi^i \widehat{\bar X}_i$. Under the infinitesimal transformation $\delta \Omega_g$ in (\ref{gauge_infinitesimal}), the metric transforms as
\eq{\delta \left( \diff s^2\right) = 2 \Omega^i \eta_{ij} \Ad^*_{g^{-1}}\left(\diff \xi^i  \right).}
\subsection{Asymptotic symmetries}
A \emph{symmetry} in the present context refers to a (gauge) transformation that leaves the metric (and the MC form) invariant, i.e. an isometry. An \emph{asymptotic symmetry} refers to a gauge transformation that leaves the metric asymptotically invariant at the boundary, here in the limit $r \to \infty$. In addition we here demand that the gauge field variation $\delta \Omega_g$ also vanishes at the boundary, which corresponds to a rather strict fall off condition for the metric. In addition we demand that the corresponding vector fields tend to a finite limit on the boundary. Once such an asymptotic symmetry has been found, one can use the above formulae to find e.g. $\widetilde{\Omega}_g$ by integrating the orbits, which is then a general expression for an asymptotically $AdS_3$ gauge field.\\

The equation for an asymptotic symmetry can now be expressed concisely as
\eq{\Ad^{-1}_{g} \cdot \diff \xi \sim 0, \label{asymptotic.symmetry.equation}}
where the "$\sim$" sign simply means asymptotically zero. The polar $k h k'$ decomposition can be used also in the adjoint representation to write $\Ad_{g} = \Ad_k \Ad_h \Ad_{k'}$. The solution for the asymptotic symmetries now proceeds in three steps. The first step is to solve the linear algebraic problem  \ref{asymptotic.symmetry.equation} asymptotically.By noting that $\det \Ad_{k}=1$ and that
\eq{\Ad^{-1}_h = \left( \begin{matrix}
\cosh r & 0 & \sinh r \\ 
0 & 1 & 0 \\
\sinh r & 0 & \cosh r
\end{matrix}\right) = \frac{1}{2}\left( \begin{matrix}
e^r & 0 & e^r \\ 
0 & 2 & 0 \\
e^r & 0 & e^r
\end{matrix}\right) + \mathcal O (e^{-r}),}
a non-trivial asymptotic solution can be obtained by solving
\eq{\Ad^{-1}_h \Ad^{-1}_{k} \cdot \diff \xi \doteq \Ad^{-1}_h \diff \tilde \xi \sim 0,}
which yields
\eq{\diff \tilde \xi = \left( \begin{matrix}
-1\\
0\\
1
\end{matrix}\right) w (u,v)}
where $w (u,v)$ is a so far arbitrary one form on $G$. Note that we are assuming that the symmetries approach a finite limit at the boundary, and that therefore to leading order there is no radial dependence in $\diff \xi$. We have
\eq{Ad_k = \left( \begin{matrix}
\cos u & \sin u & 0 \\ 
-\sin u & \cos u & 0 \\
0 & 0 & 1
\end{matrix}\right)}
which gives finally
\eq{\diff \xi = \left( \begin{matrix}
-\cos u\\
\sin u\\
1
\end{matrix}\right) w (u,v).}
The second steps involves applying the integrability condition $\diff^2 \xi = 0$, which can easily be seen to result in $w(u,v) = w'(u) \diff u$ for some function $w'(u)$. The third step is to integrate the $\diff \xi$. Denoting $w(u) \doteq \alpha(u) + \alpha''(u)$ the solution can then be written
\eq{\xi = \left( \begin{matrix}
-\alpha'(u)\sin u  -\alpha''(u)\cos u  \\ 
-\alpha'(u)\cos u  + \alpha''(u)\sin u  \\
\alpha(u) + \alpha''(u)
\end{matrix}\right),}
which results in the vector field
\eq{\widehat V \doteq \xi^i \widehat{\bar X}_i = -\alpha'(u) \partial_r + \alpha(u) \partial_u+2 e^{-r} \alpha''(u)\partial_v.}
Expansion in e.g. a basis $\alpha (u) = u^{n+1}$ would produce the Witt algebra.\\

The orbits of the above vector fields, parametrized by $s$ and by denoting $\widetilde u \doteq u(1)$ and $\alpha\! \left(\widetilde u \right) \doteq \alpha(u) e^{\varphi (u)}$, result in the finite transformations (up to order $\sim e^{-r}$)
\eq{\left\{ \begin{aligned}
\widetilde u &= \textstyle{\int\limits}^u\!\! \du' e^{\varphi(u')}\\
\widetilde r &= r-\varphi(u) \\
\widetilde v &= v+2 e^{-r}\varphi'(u).
\end{aligned}
\right. }

\subsection{Asymptotically $AdS_3$ metrics and Energy-momentum tensor}
Denoting $\tilde g = e^{\tilde u X_3}e^{\tilde r X_2}e^{\tilde v X_3}$, we have
\eq{\Omega_{\tilde g} = {\tilde g}^{-1} \diff \tilde g \doteq \widetilde \Omega^i X_i}
with
\eq{\left\{ \begin{aligned}
\widetilde \Omega^1 &= \Omega^1 +e^{-r} \left( 2 \cos v \varphi'(u) \dr +\frac{1}{2}\cos v \left[ 2\varphi'(u)^2 -e^{2\varphi (u)} +1\right]\du \right)\\
\widetilde \Omega^2 &= \Omega^2 +e^{-r} \left( 2 \sin v \varphi'(u) \dr +\frac{1}{2}\sin v \left[ 2\varphi'(u)^2 -e^{2\varphi (u)} +1\right]\du \right)\\
\widetilde \Omega^3 &= \Omega^3 +e^{-r} \left( 2 \varphi'(u) \dr +\frac{1}{2} \left[4\varphi''(u)+e^{2\varphi (u)} -1\right]\du \right)
\end{aligned}
\right. }
to order $\sim e^{-r}$. The corresponding metric is
\eq{\widetilde{\diff s^2} \doteq 2 \tr \left( \Omega_{\tilde g}\cdot \Omega_{\tilde g}\right)= \diff s^2 + T(u) \du^2 + \mathcal O \left( e^{-r}\right),\label{asymptotic.metric}}
where
\eq{T(u) \doteq -2 \varphi ''(u)+\varphi '(u)^2-e^{2 \varphi (u)}+1\label{energy-momentum}}
can be recognized as the holomorphic/ left moving component of the energy-momentum tensor of the two dimensional Liouville like conformal field theory\cite{NavarroSalas}. A corresponding result can be obtained with right actions.

\section{Concluding remarks}
While the results of the above analysis are not new, the Lie group theoretic techniques by which they were obtained are. Two features of the technique deserve special comment. Firstly, the equations (\ref{asymptotic.symmetry.equation}) to be solved are much simpler than the asymptotic Killing equations. Secondly, the fall-off conditions of an asymptotically $AdS_3$ metric is enforced by the vanishing of the variation of the gauge field at the boundary. Furthermore, the idea is of course to apply these techniques in the cases $SL(N)$ with $N>2$ and the corresponding symmetric spaces. Indeed, it has already been shown in \cite{arponen} that an infinite number of asymptotic symmetries exists on the boundaries of the symmetric space $SL(3)/SO(3)$, but the analysis did not extend beyond the leading order, and therefore no "anomaly" similar to as in eq. (\ref{asymptotic.metric}) was obtained. The reason for this is simply that the metric approach is too cumbersome and will lead to obscure and very difficult equations to be solved. The purpose of the present analysis is to enable the discovery of next to leading order asymptotic symmetries and to find the corresponding anomaly term.

\bibliographystyle{plain}
\bibliography{bibsit3}
\end{document}